\documentclass[aps,prd,twocolumn,groupedaddress,showpacs,floatfix]{revtex4}

\usepackage{graphics}
\newcommand{\g}{g_*}
\newcommand{\gT}{g_*(T)}
\newcommand{\Q}{T_c}

\newcommand{\be}{\begin{equation}}
\newcommand{\ee}{\end{equation}}
\begin{document}

\title{Effective degrees of freedom during the radiation era}
\author{Thomas S. Coleman}\email[]{tom@hilerun.org}
\affiliation{15 Grigg St., Greenwich, CT 06830, USA}

\author{Matts Roos}\email[]{Matts.Roos@helsinki.fi}
\affiliation{Department of Physical Sciences, Division of High Energy Physics,
University of Helsinki, Helsinki}

\date{\today}

\begin{abstract}
We update the curves of the effective degrees of freedom for
energy density $g_*(T)$ and for the entropy density $g_{*S}(T)$
during the era of radiation domination in the Universe. We find
that a plain count of effective degrees of freedom sets an upper
limit to the temperature of the quark-hadron transition at $\Q<
235$ MeV for the energy density and $\Q< 245$ MeV for the entropy
density.
\end{abstract}
 \maketitle
As is well known, the thermal history of the Universe during the
radiation era depends on the effective degrees of freedom
$g_{eff}(T)$ or $\gT$ contributed by the particles in thermal
equilibrium at temperature $T$. The function $\gT$ was first
computed over the eV to TeV range in 1981 \cite{Olive} and later
updated in 1988 \cite{Srednicki} and 1990 \cite{Kolb}, and has
often been quoted without updating since then ({\it e.g.}
\cite{Sarkar}, \cite{Roos}). Over time our knowledge of the
properties of the particles determining this function have
undergone various modifications, particularly at high temperatures
due to improved values for the top quark mass, improved limits for
the Higgs boson, and improved detail on the hadron spectrum
\cite{Hagiwara}. This calls for an updating of $\gT$.

Recall that the light nuclei are synthesized at temperatures $T\ll
m_e$ when all particles have decoupled and $\g$ receives
contributions only from the three relativistic neutrino species
and the photon. Each neutrino species then contributes 7/4 degrees
of freedom, corrected for the evolution of the neutrino
temperature through the period of $e^{\pm}$ annihilation, and the
photon contributes $g_{\gamma}=2$. The function $\gT$ then has the
low-energy limit

\be \g=g_{\gamma}+3\ \frac{7}{4}\left(\frac{4}{11}\right)^{4/3}
\approx 3.363\ . \ee

At temperatures above this low-energy limit but in the `confined'
phase below the quark-hadron phase transition $\Q$, the hadrons
contribute by the pions, $g_{\pi} = 3$, and by the low-energy
tails of the heavier mesons and \hbox{baryons}. At all temperatures the
three charged \hbox{leptons} contri\-bute by $g_{\ell}=
3\times\frac{7}{2}$, and the vector bosons by $g_{W^{\pm}}= 6$ and
$g_{Z^0}=3$. In the `deconfined' phase above $\Q$ the hadron
contributions are replaced by the six \hbox{quarks}, each of which
contribute $g_q=12$, and by the gluons, $g_g=16$, and the Higgs
boson, $g_{H^0}=1$. We take their masses from the Review of
Particle Physics \cite{Hagiwara} which, for the $u,\ d$ and $s$
quarks implies current-quark masses.

The total energy density of all species in equilibrium at photon
temperature $T$ can be written \cite{Kolb}

\be \rho =\frac{\pi^2}{30}\g T^4\ .\ee

\noindent Summing over all species $i$ with mass $m_i$, the
effective degrees of freedom $\g$ for the energy density is given
\cite{Kolb} by

\be \g = \sum_i\left(\frac{T_i}{T}\right)^4
g_i\frac{15}{\pi^4}\int_{x_i}^{\infty}
{\frac{(u^2-x_i^2)^{1/2}u^2}{e^u\pm 1}du}\ .\ee

\noindent Here $T_i$ is the temperature of species $i$ if it has a
thermal distribution with a different temperature than photons
({\it e.g.} neutrinos), $x_i\equiv m_i/T_i$, and the $\pm$ sign
implies + for fermions and -- for bosons. A difference from Kolb
and Turner \cite{Kolb} is that we neglect the chemical potential
which one can do for a thermal radiation background.

The entropy density is defined by

\be s=\frac{2\pi^2}{45}g_{*S}T^3\ ,\ee

\noindent where $g_{*S}$ is the effective degrees of freedom . The
contribution from all species $i$ to the entropy density is then
\cite{Kolb}

$$
 g_{*S}=\sum_i\left(\frac{T_i}{T}\right)^3\frac{15
g_i}{4\pi^4}\ (3\int_{x_i}^{\infty}{\frac{\sqrt{u^2-x_i^2}\
u^2}{e^u\pm
 1}}du\ + $$
\be \int_{x_i}^{\infty} {\frac{\sqrt{u^2-x_i^2}^{\ 3}}{e^u\pm
1}}du\ )\ .\ee

These integrals can be evaluated numerically once one knows which
particle species should be included in the sums, and what value to
take for the transition temperature $\Q$.

To answer the first question, consider the thermalization of the
$\rho(750)$ meson (in units of $\hbar=c=k=1$). The $\rho(750)$
number density at temperature $T=0.2$ GeV is

\be N_{\rho}=n_{spin}(2\pi m_{\rho}T)^{3/2}e^{-m_{\rho}/T}\approx
0.065\ \hbox{GeV}^3\ .\ee

\noindent The cms velocity in elastic
$\pi+\rho\rightarrow\pi+\rho$ collisions is $v_{cms}\approx 0.15$.
Taking the cross-section to be about 30 mb or $\sigma\approx
7.7\times 10^7\ \hbox{GeV}^{-2}$ we find the reaction rate

\be \Gamma=\langle N_{\rho} v_{cms}\sigma\rangle\approx 7.5\times
10^5\ \hbox{GeV}\ .\ee

\noindent This should be compared to the decay rate of the rho
meson, $\Gamma_{\rho}=0.15$ GeV. Thus there is plenty of time for
rho mesons in the tail of their thermal distribution to thermalize
in collisions against pions.

One comes to the same conclusion when one studies the reaction
$\gamma+\rho\rightarrow\gamma+\rho$ : the reaction rate is then
$1.7\times 10^4$ GeV for an approximate cross-section of 0.1 mb.

One also comes to the same conclusion when one turns to heavier
meson resonances and to baryon resonances. We therefore include
all known meson and baryon resonances \cite{Hagiwara} up to 3 GeV
mass in the sums in $\g$ and $g_{*S}$ at temperatures below $\Q$.
This is the procedure followed also by the earlier studies
\cite{Olive,Srednicki,Kolb,Kolb2}.

In spite of the strong exponential decrease in the number density
of non-relativistic species with heavy masses $m_i>\Q$, such
species will contribute significantly below $\Q$. For example the
proton and neutron contribute 0.25 each to $\gT$ at $T=150$ MeV
(versus an unattenuated 3.5) even though their mass of roughly 939
MeV is 6.3 times higher. The sum of unattenuated mesonic and
baryonic degrees of freedom below 3 GeV mass is $g_{mesons}=495$
and $g_{baryons}=320$, respectively. As a result $\gT$ rises quite
rapidly for temperatures above $T=100$ MeV.

Near the transition temperature $\Q$ the ideal gas approximation
breaks down and the transition from the deconfined quark phase to
the confined hadronic phase can only be studied with finite
temperature lattice simulations. No reliable value for $\Q$ has
yet been obtained, but it is believed to be near 170 MeV
\cite{Laine}. Srednicki et al. \cite{Srednicki} proposed that the
transition my be smooth and bracketed by the cases $\Q=150$ MeV
and 400 MeV. Their procedure to obtain smooth curves in the
context of the `bag' model has only been elucidated by Sarkar in a
footnote on p. 1513 \cite{Sarkar} which refers to a private
communication from K. A. Olive. But whether the change is smooth
or discontinuous still remains open \cite{Laine}.

\begin{figure}
\includegraphics{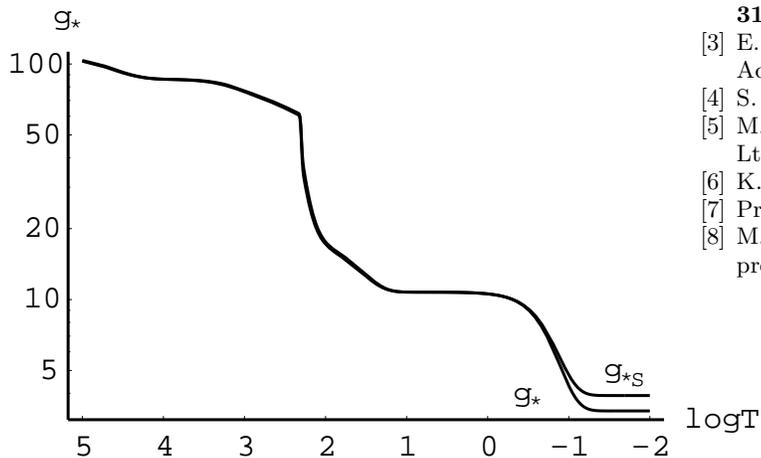}
\caption{\label{f1} Effective degrees of
freedom for the energy density $\g(T)$ and for the entropy density $g_{*S}(T)$.
 The quark-hadron transition is chosen to occur at
200 MeV.}
\end{figure}

In Fig.~\ref{f1} we plot $\gT$ and $g_{*S}(T)$ as (almost)
discontinuous functions of $\log T$ for the range 100 GeV $>T>$ 10
keV. For purposes of the figure we have somewhat arbitrarily
chosen the discontinuity at $\Q=200$ MeV (as in \cite{Kolb}) and a
smoothing naive transition function that mixes the quark and
hadronic phases over a range of roughly 5 MeV on either side of
$\Q$. Since it is quite laborious to calculate the curves in Fig.~
\ref{f1} and difficult to read them precisely there, we give here
some numerical approximations over the range $T = 155-300$ MeV.
For the hadronic phase we have

$$ \gT = 2.388\times 10^{-4}\ T^{2.287},\ \ \ \hbox{max error}\
5.3\% \ ,$$
\vspace{-0.8cm}
\be g_{*S}(T) = 3.631\times 10^{-4}\ T^{2.192},\ \ \ \hbox{max error}\
5.5\%\ .\ee

\noindent For the quark phase we have

$$ \gT = 24.604 \times T^{0.1713},\ \ \ \hbox{max error}\ 0.2\% \
,$$
\vspace{-0.8cm}
\be g_{*S}(T) = 22.961 \times T^{0.1810},\ \ \ \hbox{max error}\ 0.2\%
\ .\ee

\begin{figure}
\includegraphics{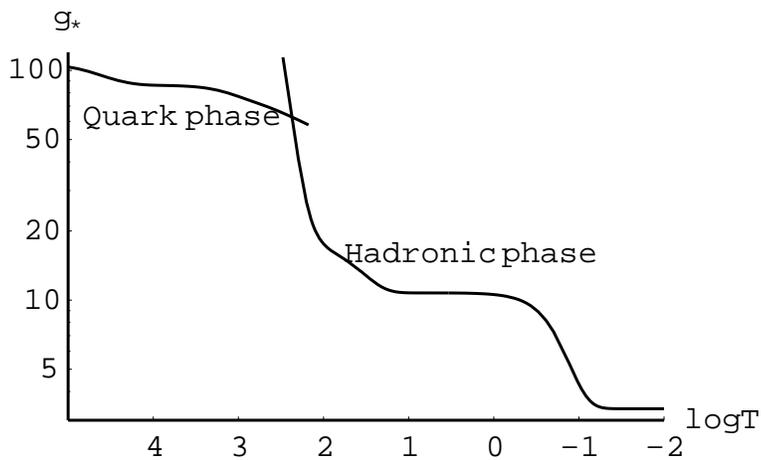}
\caption{\label{f2} Effective degrees of
freedom $\g(T)$ due to the quark and hadronic phases,
respectively. For the quark phase we have taken $\Q= 150$ MeV, for
the hadronic phase $\Q= 300$ MeV. The cross-over occurs at 235
MeV}
\end{figure}

It is instructive to exhibit $\gT$ in the quark phase and the
hadronic phase separately, as we have done in Fig.~\ref{f2}. This
shows a very well defined cross-over at 235 MeV. Taking this
temperature to be $\Q$ one obtains a curve which is
indistinguishable from the curves in Fig.~\ref{f1} (because of the
logarithmic scales). The cross-over temperature is defined quite
precisely due to the steepness of the hadronic phase, but it does
of course depend to some extent on the precision of the hadronic
resonance energies, as well as on unknown particles and
interactions above 1 TeV.

Our most interesting conclusion is that a plain count of effective
degrees of freedom sets an upper limit to the temperature of the
quark-hadron transition. From the $\gT$ cross-over one has

\be \Q\lesssim 235\ \hbox{MeV}\ ,\ee

\noindent and from the  $g_{*S}(T)$ cross-over

\be \Q\lesssim 245\ \hbox{MeV}\ .\ee

Higher $\Q$ values would give an unphysical spike and lower values
would increase the discontinuous drop in $\gT$ just below $\Q$.


\begin{acknowledgments}
We thank Kari Enqvist and Rocky Kolb for useful discussions.
\end{acknowledgments}

\end{document}